\documentclass[bjps]{imsart}

\RequirePackage{amsthm,amsmath,amsfonts,amssymb}
\RequirePackage[authoryear]{natbib}
\RequirePackage[colorlinks,citecolor=blue,urlcolor=blue]{hyperref}
\RequirePackage{graphicx}

\usepackage{verbatim}
\usepackage{amsmath}
\usepackage{graphicx,psfrag,epsf}
\usepackage{enumerate}
\usepackage{natbib}
\usepackage{multirow}
\usepackage{epstopdf}
\usepackage{array}
\usepackage{url} 
\usepackage{scalefnt} 
\usepackage[shortlabels]{enumitem} 

\usepackage{makeidx}     
\usepackage{multicol}
\usepackage{amssymb}
\usepackage{latexsym}
\usepackage{amsmath}
\usepackage{amsfonts}
\usepackage{amssymb}
\usepackage{amsopn}
\usepackage{amsthm}
\usepackage{multirow}
\usepackage{scalefnt}
\usepackage{natbib}
\usepackage{booktabs}
\usepackage{rotating}
\usepackage[usenames]{color} 

\usepackage{graphicx,psfrag,epsf}
\usepackage{enumerate}
\usepackage{epstopdf}
\usepackage{array}
\usepackage{url} 
\usepackage[shortlabels]{enumitem} 

\startlocaldefs
\theoremstyle{plain}

\theoremstyle{remark}

\newcommand{\mat}[1]{\mbox{\boldmath{$#1$}}}

\numberwithin{equation}{section}

\endlocaldefs

\begin{document}






\begin{frontmatter}
\title{Estimation and variable selection in joint mean and dispersion models applied to mixture experiments}

\runtitle{Estimation and variable selection in JMMD applied to mixture experiments}

\begin{aug}
\author[A]{\inits{E. R.}\fnms{Edmilson Rodrigues} \snm{Pinto}\ead[label=e1]{edmilson.pinto@ufu.br}}\thanksref{t1}
\and
\author[A]{\inits{L. A.}\fnms{Leandro Alves} \snm{Pereira}\ead[label=e2,mark]{leandro.ap@ufu.br}}\thanksref{t1}

\thankstext{t1}{The authors thank Fapemig for financial support.}

\runauthor{Pinto, E.R. and Pereira, L.A.}

\address[A]{Federal University of Uberlândia, Uberlândia, Brazil,
\printead{e1,e2}}
\end{aug}

\begin{abstract}
In industrial experiments, controlling variability is of paramount importance to ensure product quality. Classical regression models for mixture experiments are widely used in industry, however, when the assumption of constant variance is not satisfied, the building of procedures that allow minimizing the variability becomes necessary and other methods of statistical modeling should be considered. In this article, we use the class of generalized linear models (GLMs) to build statistical models in mixture experiments. The GLMs class is general and very flexible, generalizing some of the most important probability distributions, and allows modeling the variability through the methodology of the joint modeling of mean and dispersion (JMMD). This paper shows how the JMMD can be used to obtain models for mean and variance in mixture experiments. We give a comprehensive understanding of the procedures for estimating parameters and selecting variables in the JMMD. The variable selection procedure was adapted for the case of mixture experiments, where the verification of constant dispersion is ensured by the existence of only the constant term in the dispersion model; the absence of the constant term or the existence of any other term in the dispersion model implies non-constant dispersion. A simulation study, considering the most common case of Normal distribution, was used to verify the effectiveness of the proposed variable selection procedure. A practical example from the Food Industry was used to illustrate the proposed methodology. 

\end{abstract}

\begin{keyword}
\kwd{Variance modeling}
\kwd{process and noise variables}
\kwd{generalized linear models}
\kwd{model selection}
\kwd{robust design experiment}
\end{keyword}

\end{frontmatter}

\section{Introduction}
\label{introd} 

Experiments with mixtures involve the mixing or blending of two or more ingredients to form an end product. For this type of experiment is of interest to determine the proportions of the mixture components which lead to desirable results with respect to some quality characteristic of interest. An excellent introduction to mixture experiments, including a chronological sequence of papers that appeared in the statistical literature in this area from 1953 to 2009, is given by \cite{Cornell2}.

Classical regression models are usually used in mixture experiments, which assume normality of errors, constant variance and linearity of systematic effects. However, situations may arise where such assumptions cannot be completely satisfied. According to \cite{NelderLee}, for many  data sets the assumptions underlying classical linear models are unsatisfactory and transformations of response variable cannot necessarily satisfy all three criteria including the normality. 

Generalized linear models (GLMs) allow the analysis of data for which the assumptions of the classical regression model are not satisfied and provide a general framework for modeling variability through joint modeling of mean and dispersion (JMMD) that allows joint models to be build for the mean and dispersion. The JMMD was introduced by \cite{NelderLee} as an alternative to Taguchi's methods in quality-improvement experiment and provides a methodology to find and check the fit of the models found with a solid statistical basis. Further examples of applications appeared in \cite{LeeNelderI} and \cite{LeeNelderII}. In generalized linear models, the response variance is equal to the dispersion parameter multiplied by a function of the mean, called the variance function. Only in the case of normal distribution, where the variance function is the identity function, the dispersion parameter coincides with the variance of the response, see \cite{McCullaghNelder} for more details. 

In mixture experiments, modeling of variance is only considered in situations where noise variables exist. Noise variables or uncontrollable variables are process variables that vary naturally and uncontrollably during the production but may be controlled during the experiment, such as humidity or temperature, see \cite{Montgomery}. p. 571. Process variables are factors in an experiment that do not form any portion of the mixture but whose levels when changed could affect the blending properties of the ingredients, see, for example, \cite{Cornell2}, p. 247, for more details.

Statistical modeling of variance in mixture experiments with noise variables has been considered in \cite{SteinerHamada}, who proposed a combined mixture-process-noise variable model, built and solved an optimization problem to minimize a quadratic loss function, taking into account both the mean and variance of response. Another approach to modeling the variance in mixture experiments with noise variables is due to \cite{Goldfarb} using the delta method, which employs a first-order Taylor series approximation of the regression model at a vector of noise variables. The delta method is a well-known technique, based on Taylor series expansions, for finding approximations to the mean and variance of functions of random variables. For a comprehensive treatment of the delta method, see \cite{CasellaBerger}, p. 240. An application of the delta method for modeling variability in mixture experiments with noise variables in food engineering is also presented in \cite{GranatoCaladoPinto}. However, in all approaches presented for modeling the variance, normality of errors and constant dispersion parameter were considered. Thus, the mean model was considered normal homoscedastic and the modeling of variance was due only to noise variables.

The approach we consider is quite general because, when considering the class of generalized linear models, distributions other than the normal distribution may be considered. In addition, considering a dispersion model, the variability modeling process allows modeling the variance not only in situations where noise variables exist.

The purpose of this article is to show how joint modeling of mean and dispersion can be applied to model the variability in mixture experiments. We give a comprehensive understanding of the parameter estimation process in JMMD and we propose a procedure of variable selection adapted to the conditions of the mixture experiments. A simulation study was used to exemplify the proposed variable selection procedure. The considered methodology was applied to an example from the Food Industry, where a detailed explanation of the proposed method was presented.

The paper is organized as follows. In Section \ref{MixtureExperiments}, we introduce the mixture experiments. In Section \ref{JMMD}, we provide a summary of the joint modeling of mean and dispersion, discussing the principal points of the theory. A variable selection process, adapted to mixture experiments, is also presented. A simulation study, considering Normal distribution for the response variable, is shown in Section  \ref{NumericalEvaluation}. In Section \ref{Application}, we show how the joint modeling of mean and dispersion can be applied to a practical bread-making problem involving three mixture ingredients, three types of flour,  and subject to two noise variables, mixing time and proofing time for the dough. Final considerations are given in Section \ref{Conclusion}.

\section{Mixture experiments} 
\label{MixtureExperiments}

A mixture experiment involves mixing proportions of two or more components to make different compositions of an end product. In mixture experiments, the main aim is at determining the proportions of the mixture components which lead to desirable properties of the response variable. Mixture component proportions $x_i$ are subject to the constraints

\begin{equation} \label{eq1}
0\le x_i \le 1 \quad \quad i=1,2,\dots,a \quad \quad \textrm{and} \quad \quad \sum_{i=1}^{a} x_{i}=1,
\end{equation}

\noindent where $a$ is the number of components involved in the mixture experiment. Consequently, the design space or experimental region is a $(a-1)$-dimensional simplex. A $k$-dimensional simplex experimental region is a $k$-dimensional regular convex polytope determined by the mixture restrictions. The experimental region is part of a $k$-dimensional simplex if there are mixture constraints, i.e., if there are further conditions on the proportions such as $c_i \le x_i \le d_i$ for $i=1,2, \dots, a-1$, with the proportion $x_a$ taking values which make up the mixture. The values of $c_i$ and $d_i$ are, respectively,  the lower and upper limits of mixture. If the mixture constraints are consistent, see \cite{Piepel1}, the design space is an irregular convex polytope.

In consonance with \cite{Khuri}, consider the standard model in a mixture experiment the linear model  $\mat{Y}=\mat{X}\mat{\beta} + \mat{\epsilon}$, where $\mat{Y}^t=(Y_1,\ldots,Y_n)$ is a response vector, $\mat{X}$ is a known design matrix of order $n \times n_p$ and rank $n_p \le n$, $\mat{\beta}^t=(\beta_1, \ldots, \beta_{n_p})$ is an unknown parameter vector, and $\mat{\epsilon}^t=(\epsilon_1, \ldots \epsilon_n)$ is a random error vector with a zero mean and a variance-covariance matrix $\sigma^2 \mat{I}$, with $\sigma^2=Var(\epsilon_i)$, for $i=1,\ldots,n$, and $\mat{I}$ is the is the identity matrix of order $n$. The response value at any point $\mat{x}^t=(x_1, \ldots, x_a)$ in a region of interest $\cal{R}$ is given by $Y(\mat{x})= \zeta(\mat{x}, \mat{\beta})+\epsilon$, where the linear predictor $\zeta(\mat{x},\mat{\beta})=\mat{f}^t(\mat{x})\mat{\beta}$ is a polynomial function of a certain degree in $x_1, \ldots, x_a$. The vector $\mat{f}^t\left(\mat{x})=(f_1(\mat{x}), \ldots, f_{n_p}(\mat{x})\right) $ is of the same form as a row of $\mat{X}$, but is evaluated at the
point $\mat{x}$. We can see that the linear predictor $\zeta(\mat{x}, \mat{\beta})=E(Y(\mat{x}))$.

Models commonly used in experiments with mixture are the so-called Scheff\'{e}'s models. \cite{Scheffe1} proposed canonical polynomial models for mixture experiments, considering a reparametrization of the polynomial regression model, based on the mixture constraint in equation (\ref{eq1}). For a comprehensive understanding of Scheff\'{e}'s models, see \cite{Cornell} or \cite{Cornell2}. The linear predictor of Scheff\'{e}'s first-order polynomial model is given by

\begin{equation} \label{eq33}
 \zeta(\mat{x}, \mat{\beta})=\sum_{i=1}^{a} \beta_{i}x_{i},
\end{equation}

\noindent which does not explicitly contain the constant term. The linear predictor of Scheff\'{e}'s second-order model (quadratic model) is given by

\begin{equation} \label{eq4}
\zeta(\mat{x}, \mat{\beta})=\sum_{i=1}^{a} \beta_{i}x_{i}+ \sum_{i=1}^{a-1} \sum_{j=i+1}^{a} \beta_{ij}x_{i}x_{j}.
\end{equation}

For higher order models, the reparametrization does not lead to simple models; for example, for the third order model (cubic model), the linear predictor of Scheff\'{e}'s canonical polynomial model is given by

\begin{eqnarray} \label{eq5}
\zeta(\mat{x}, \mat{\beta}) & = & \sum_{i=1}^{a} \beta_{i}x_{i}+ \sum_{i=1}^{a-1} \sum_{j=i+1}^{a} \beta_{ij}x_{i}x_{j} + \sum_{i=1}^{a-1} \sum_{j=i+1}^{a} \beta_{i-j}x_{i}x_{j}(x_{i}-x_{j}) + \nonumber \\
                           &   & \sum_{i=1}^{a-2} \sum_{j=i+1}^{a-1} \sum_{k=j+1}^{a} \beta_{ijk}x_{i}x_{j}x_{k}.
\end{eqnarray} 

For estimation of model parameters, Scheff\'{e} proposed the simplex-lattice designs ($a,m$), which are characterized by symmetrical arrangements of points within the experimental region, where $m$ is the number of equally spaced parts in the interval $(0,1)$. The number of parameters in the canonical polynomial model is exactly equal to the number of points chosen within the experimental region. The simplex-lattice design ($a, m$) consists of $\frac{(a + m-1)!}{m!(a-1)!}$ experimental points. Each component of the mixture takes $m + 1$ values equally spaced between 0 and 1, that is, $x_{i} = 0, \frac{1}{m}, \frac{2}{m}, \ldots, \frac{m}{m} = 1$ for $i = 1, \ldots, a$. All mixtures involving these proportions are used in the experimental design. For more details on simplex-lattice designs, see \cite{Cornell}.

Alternatively to the Scheff\'{e}'s canonical models, we can also use models of mixture with slack variable. Due to the mixture restrictions (\ref{eq1}), the proportions of the components are not independent; thus, knowing the proportions of the first $a-1$ components we can determine the proportion of the remaining component. In this way, the mixture models with slack variable can be obtained from the Scheff\'{e}'s canonical models by replacing the slack variable chosen, say $x_l$, by $1-\sum_{i \not=l}^{a} x_{i}$.  The purpose of this process is to produce mixture models that depend on $a-1$ independent variables. For example, by defining $x_a$ as the slack variable, the linear predictor for the mixture model with slack variable obtained from Scheff\'{e}'s quadratic model (\ref{eq4}) is given by

\begin{equation} \label{eq7}
\psi(\mat{x}, \mat{\alpha})= \alpha_0 + \sum_{i=1}^{a-1} \alpha_{i}x_{i} + \sum_{i=1}^{a-1}\alpha_{ii}x_{i}^{2} + \sum_{i=1}^{a-2} \sum_{j=i+1}^{a-1} \alpha_{ij}x_{i}x_{j}.
\end{equation} 
 
Note that, by comparing the coefficients of models (\ref{eq4}) and (\ref{eq7}), we have the following equivalences $\alpha_0=\beta_a$, $\alpha_{i}=\beta_{i}-\beta_{a} +\beta_{ia}$, $\alpha_{ii}= -\beta_{ia}$ and $\alpha_{ij}=\beta_{ij} -(\beta_{ia} + \beta_{ja})$, with $i=1,\ldots, a-1$ and $j=2,\ldots,a-1$.

The use of models with slack variable as well as the choice of which component should be designated as a slack variable is not clearly presented in the literature. \cite{Cornell2} highlights that the choice of which of the $a$ components to designate as the slack variable has not been defended from either a theoretical or a practical point of view. A discussion of the pros and cons of using slack variable is given by \cite{Cornell1}; see also \cite{Smith}. Comparative studies between models of mixtures with slack variable and Scheff\'{e}'s models are given in \cite{Cornell}, \cite{CornellGorman} and \cite{Khuri}.

When in a mixture experiment, the response depends not only on the proportion of the mixture components, but also on the processing conditions, such as temperature, humidity or heating time, we have a mixture experiment with process variables. Problems of mixture experiments with process variables arise when in the mixture experiment the property of interest is a function of the proportions of the ingredients and of other factors that do not form any portion of the mixture. These factors, which depend on process conditions, are called process variables, see \cite{Cornell}, p. 354.

Usually, in problems of mixture experiments including process variables, the goal is determining the proportions of the mixture components along with situations of process conditions, in order to achieve an optimal or desirable property on the response variable.

Thus, if in addition to the $a$ mixture components $\mat{x}^t=(x_1,\dots,x_{a})$ there are $r$ process variables $\mat{z}^t=(z_1,\dots,z_r)$,  we can consider additive models like $\eta(\mat{x},\mat{z}, \mat{\beta}, \mat{\gamma})=\zeta(\mat{x}, \mat{\beta})+\vartheta(\mat{z}, \mat{\gamma})$ or complete cross product models of the type $\eta(\mat{x},\mat{z}, \mat{\beta}, \mat{\gamma})=\zeta(\mat{x}, \mat{\beta})\vartheta(\mat{z}, \mat{\gamma})$ or combinations of these such as $\eta(\mat{x},\mat{z}, \mat{\beta}, \mat{\gamma})=\zeta(\mat{x}, \mat{\beta})+ \nu(\mat{x},\mat{z}, \mat{\beta}, \mat{\gamma})$, where $\vartheta(\mat{z}, \mat{\gamma})$ represents the linear predictor for the process variable model and $\nu(\mat{x},\mat{z}, \mat{\beta}, \mat{\gamma})$ comprises products of terms in $\zeta(\mat{x}, \mat{\beta})$ and $\vartheta(\mat{z}, \mat{\gamma})$. In general, the methodology used to construct mixture designs involving process variables is a combination of two design spaces, i.e.,  one design space being the ($a-1$)-dimensional simplex, while the other, the design space for the process variables, is the $[-1,1]^r$ hypercube. Factorial and fractional factorial designs are a subset of possible designs on this hypercube. For more details on mixture experiments with process variables, see \cite{Cornell}.

Process variables can be controlled during the experiment, however some process variables can be difficult to control or uncontrollable during the production process, in which case they are treated as noise variables. Noise variables are considered random variables with a supposedly known probability distribution. Thus, the presence of noise variables in the model means that the variance can no longer be assumed to be constant. One way to control the variability present in the model is by modeling the variance, see \cite{Goldfarb} and \cite{GranatoCaladoPinto}. 

A comprehensive presentation of several aspects of statistical modeling and experimental design involving mixture experiments is given by \cite{Cornell}. Practical aspects of the theory are presented by \cite{Smith}. \cite{Piepel} makes a bibliographical survey on mixture experiments over a period of 50 years, ranging from 1955 to 2004; see also \cite{Cornell2}. A bibliography of papers on mixture designs with process variables as well as various other types of mixture experiments is given by \cite{PiepelCornell}. 

\section{Joint modeling of mean and dispersion}
\label{JMMD}

The class of generalized linear models, introduced by \cite{NelderWedderburn}, is a general framework for handling a range of common statistical models for normal and non-normal data. Generalized linear models provide a straightforward way of modeling non-normal data when the usual regression assumptions are not satisfied \citep{Jorgensen}.

As pointed out by \cite{LeeNelderII}, GLMs extend classical linear models in two ways: (i) the assumption of the normal distribution to describe the random part of the model is extended to one-parameter exponential families, and (ii) additivity of  the effects of explanatory variables is assumed to hold on some  transformed scale defined by a monotone function called the link  function. A generalized linear model splits the finding of an additive scale for the effects of the explanatory variables from  the specification of the error structure. The scale on which the effects are assumed additive is related to  the mean of the error distribution by the link function. Thus we write $\eta$ for the linear part of  the model, where $\eta$ is called the linear predictor, and connect this with the mean, by the link function  $\eta = g(\mu)$. We do not transform the data to produce  additivity; rather we transform the hypothetical mean  value. $\eta=\eta(\mat{x},\mat{\beta})$ is a function of unknown parameters $\mat{\beta}$ and the explanatory variables $\mat{x}$, which is expressed as linear combinations of the parameters $\mat{\beta}$. The variance of the response $Y$ is assumed to be proportional to the variance  function, $Var(Y) = \phi V (\mu)$. This shows that the variance splits into two parts; $\phi > 0$, called the dispersion parameter, which is independent of the mean, and $V(\mu)$, called the variance function, which describes how the variance changes with the mean.  According to \cite{Jorgensen}, the two key ingredients for a generalized linear model are the positive variance function $V(\mu)$ and the monotonic link function $g(\mu)$. Both $V(\mu)$ and $g(\mu)$ are assumed to be continuously differentiable functions of the mean $\mu$.  For example, the classical regression model is a generalized linear model with Normal distribution for the response variable $Y$, i.e. $Y \sim N(\mu, \sigma^2)$, $g(\mu)=\mu=\eta=\sum \beta_j x_j$, $\phi=\sigma^2$ and $V(\mu)=1$. For a comprehensive understanding of the theory of generalized linear models see \cite{McCullaghNelder}.

GLMs are more general than normal linear methods in that a mean-variance relationship appropriate for the data can be accommodated and an appropriate scale can be chosen for modeling the mean on which the action of the covariates is approximately linear. In generalized linear models the focus is on modeling and estimating the mean structure of the data while treating the dispersion parameter as a constant since GLMs automatically allow for dependence of the variance on the mean through the distributional assumptions. However, there are situations in which the observed data may exhibit greater variability than the one which is implied by the mean-variance relationship and thus the loss of efficiency in estimating the mean parameters, using constant dispersion models when the dispersion is varying, may be substantial \citep{AntoniadisEtal}. The joint modeling of the mean and dispersion is a general method that overcomes this difficulty, allowing the modeling of dispersion as a function of covariates, by the use of two interlinked generalized linear models.

According to \cite{NelderLee}, the method of joint modeling of mean and dispersion consists of finding joint models for the mean and dispersion. In their approach, using the extended quasi-likelihood, two interlinked generalized linear models are needed, one for the mean and the other for the dispersion. 

Let $Y_1, \ldots,Y_n$ be $n$ independent random variables with the same probability distribution, representing the dependent response variables, whose observed values are given by $y_1,\ldots,y_n$.  For the $i$th response $Y_i$ it is assumed to be known that $E(Y_i)=\mu_i$ and $Var(Y_i)=\phi_iV(\mu_i)$, where $\phi_i$ is the dispersion parameter and $V(.)$ is the variance function in GLMs. The mean and dispersion models are constructed as follows.

Suppose $\mat{t}^t=(t_1,\ldots,t_s)$ and $\mat{u}^t=(u_1,\ldots,u_v)$ are the vectors of the independent variables that affect the mean and dispersion models, respectively. The vectors $\mat{t}$ and $\mat{u}$ contain the mixture components $\mat{x}$ and additionally may also contain process variables $\mat{z}$, if they exist.

Let $\varphi$ be a link function for the mean model, i.e., for the $i$th response, $\eta_i=\varphi(\mu_i)=\mat{f}^t(\mat{t}_i)\mat{\beta}$ with $\mat{f}^t(\mat{t}_i)=(f_1(\mat{t}_i),\ldots,f_p(\mat{t}_i))$ where $f_j(\mat{t}_i)$, for $j=1,\ldots,p$, is a known function of $\mat{t}_i$ and $\mat{\beta}$ is a $p\times1$ vector of unknown parameters.

For the dispersion model is used as response variable the deviance component, which, for each observation $y_i$, is given by

\begin {equation} \label{eqdi}
d_{i} = 2 \int_{\mu_{i}}^{y_{i}}\frac{y_i- l}{V(l)}dl,   
\end{equation}

\noindent see \cite{McCullaghNelder}, p. 360. 

Following \cite{LeeNelderI}, for the dispersion model we are assuming a Gamma model with a log link function, i.e., for the $i$th response, $\xi_i=\log(\phi_i)=\mat{g}^t(\mat{u}_i)\mat{\gamma}$, with $\mat{g}^t(\mat{u}_i)=(g_1(\mat{u}_i),\ldots,g_q(\mat{u}_i))$, where $g_j(\mat{u}_i)$, for $j=1,\ldots,q$, is a known function of $\mat{u}_i$ and $\mat{\gamma}$ is a $q\times1$ vector of unknown parameters. We also define $\mat{T}=[\mat{f}(\mat{t}_1),\ldots,\mat{f}(\mat{t}_n)]^t$ the $n \times p$ design matrix for the mean model and $\mat{U}=[\mat{g}(\mat{u}_1),\ldots,\mat{g}(\mat{u}_n)]^t$ the $n \times q$ design matrix for the dispersion model.

\subsection{Estimation}
\label{Estimation}

The fitting for the JMMD uses as an optimizing criterion the extended quasi-likelihood, introduced by \cite{NelderPregibon}, see \cite{McCullaghNelder}, p. 349. For other references on extended quasi-likelihood, see \cite{PintoPereira}. In this work we use the adjusted extended quasi-likelihood, introduced by \cite{LeeNelderI} and given by

\begin{equation} \label{eqQ}
Q^{+}(\mat{\mu},\mat{\phi};\mat{y})=\sum_{i=1}^{n} -\frac{1}{2}\left( \frac{d_{i}^{*}}{\phi_{i}} + \log \{ 2\pi\phi_{i}V(y_{i})\}\right)
\end{equation}

\noindent where $d_{i}^{*}=\frac{d_{i}}{1-h_{i}}$ is the standardized deviance component and $h_{i}$ is the $i$th element of the diagonal of the matrix $\mat{H}=\mat{W}^{\frac{1}{2}}\mat{T}(\mat{T}^{t}\mat{WT})^{-1}\mat{T}\mat{W}^{\frac{1}{2}}$, being $\mat{W}$, the $n \times n$ weight matrix for the GLMs, a diagonal matrix with elements given by $w_{i}=\left(\frac{\partial\mu_{i}}{\partial\eta_{i}}\right)^{2}\frac{1}{\phi_i V(\mu_{i})}$ with $i=1,\ldots,n$. Table \ref{Tab1} shows a resume of the joint modeling of mean and dispersion. From Table \ref{Tab1}, we can observe that the standardized deviance component from the model for the mean becomes the response for the dispersion model, and the inverse of fitted values for the dispersion model provides the prior weights for the mean model.

\begin {table}[!htb]
\caption{Summary of the JMMD for the {\it i}th response}
\scalefont{0.90}
\begin{tabular}{p{4.0cm} p{3.5cm} p{4.5cm}}
\hline
Component               & Mean model                                           & Dispersion model$^{\dag}$                        \\ \hline
Response variable       &  $y_i$                                               &  $d_{i}^{*}$                                     \\
Mean                    &  $\mu_i$                                             &  $\phi_i$                                        \\ 
Variance                &  $\phi_i V(\mu_i)$                                   &  $2\phi_{i}^{2}$                                 \\
Link function           &  $\eta_i=\varphi(\mu_i)$                             &  $\xi_i=\log(\phi_i)$                             \\
Linear predictor        &  $\eta_i=\mat{f}^t(\mat{t}_i)\mat{\beta}$            &  $\xi_i=\mat{g}^t(\mat{u}_i)\mat{\gamma}$        \\
Deviance component      &  $d_i=2 \int_{\mu_i}^{y_i}\frac{y_i-l}{V(l)}dl$     &  
$2\left\{-\log\left(\frac{d_{i}^{*}}{\phi_i}\right) + \frac{(d_{i}^{*}-\phi_i)}{\phi_i}\right\}$                                   \\
Prior weight            &  $\frac{1}{\phi_i}$                                  &  $(1-h_i)/2$                                     \\ \hline
\multicolumn{3}{l}{$^{\dag}$\scriptsize{For the dispersion model we are assuming a Gamma model with logarithmic link function}}
\end{tabular}
\label{Tab1}
\end{table}

The algorithm for parameter estimation is an extension of the standard GLMs algorithm, in which the model for the mean is fitted assuming that the fitted values for the dispersion are known and that the model for dispersion is fitted using the fitted values for the mean. The fitting alternates between the mean and dispersion models until convergence is achieved. Algorithm 1 describes the parameter estimation process for joint mean and dispersion modeling. \vspace{1.5cm}

\noindent {\bf Algorithm 1}: Pseudo algorithm for joint modeling of mean and dispersion 
  
  \begin{enumerate}
   
  \item Mean model - Considering a probability distribution for the response variable, a  linear predictor and a link function, calculate $\mat{\beta}^t=(\beta_1,\ldots,\beta_p)$ using the algorithm for fitting generalized linear models (see \cite{McCullaghNelder}, p. 40).
    
 \item With the value found for $\mat{\beta}$ in step 1, calculate $\mat{d}^{*}=(d^*_1,\ldots,d^*_n)^t$.
   
 \item Dispersion model - Considering $\mat{d}^*$ as response for the dispersion model and assuming a Gamma model with logarithmic link function, calculate $\mat{\gamma}^t=(\gamma_1,\ldots,\gamma_q)$ using the algorithm for fitting generalized linear models. 
   
 \item With the value found for $\mat{\gamma}$ in step 3, calculate $\mat{\phi}^t=(\phi_1,\ldots,\phi_n)$.
   
 \item With the value found for $\mat{d}^*$ in step 2 and the value found for $\mat{\phi}$ in step 4, calculate $Q^+(\mat{\mu},\mat{\phi};\mat{y})$ and use this measure to verify the convergence of the joint model in the next iteration, that is, comparing its value with the value obtained in the next iteration. If the convergence is achieved, stop. The current $\mat{\beta}$ and $\mat{\gamma}$ are the final parameters for the joint model. Otherwise, with the value of $\mat{\phi}$, updated in step 4, calculate the new weights $w_1,\ldots,w_n$ for the mean model and go to the step 1.
 
 \end{enumerate} 

In step 5 of Algorithm 1, the new weights for the mean model in the next iteration of Algorithm 1 are  $w_{i}=\left(\frac{\partial\mu_{i}}{\partial\eta_{i}}\right)^{2}\frac{1}{\phi_i V(\mu_{i})}$ for $i=1,\ldots,n$. Note that the dispersion parameter influences the weights for the mean model only from the second iteration of the algorithm.

There are statistical packages that perform the joint modeling of the mean and dispersion. However, the Algorithm 1 can be easily implemented in software R \citep{RTeam} using the parameter estimation function for generalized linear models.

In the next subsection, we present a procedure for variable selection in JMMD applied to mixture experiments. 

\subsection{Variable selection}
\label{Variable_Selection}

The process of selecting variables considered for JMMD in mixture experiments is based on the procedure proposed by \cite{PintoPereira} for the general case of joint modeling of the mean and dispersion. 

The variable selection procedure proposed by \cite{PintoPereira} is based on hypothesis testing and the quality of the model's fit. A criterion for verifying the quality of the adjustment is used, at each iteration of the selection process, as a filter for choosing the terms that will be evaluated by a hypothesis test. The selection strategy, proposed by them and shown in Algorithm 2, consists of a three-step scheme in which the selected dispersion model is used to select the best mean model and vice versa. The selection scheme uses a recursive procedure that is only finalized when the criterion value, considered as a selection measure for the mean model, stops improving.
\vspace{0.3cm}

\noindent {\bf Algorithm 2}: Pseudo algorithm for variable selection in JMMD 

\begin{enumerate} 

\item Assuming constant dispersion, use Algorithm 3 to find the terms of the best current mean model and for this model calculate the value of the criterion considered as a selection measure.

\item Assuming that the selected current mean model is adequate, use Algorithm 3 to find the terms of the best current dispersion model.

\item Assuming that the selected current dispersion model is adequate, use Algorithm 3 to find the terms of the best current mean model and for this model calculate a new value of the criterion considered. If the updated criterion value is worse or is equal to that previously obtained, stop and the models previously found for mean and dispersion are chosen. Otherwise, return to step 2.

\end{enumerate}

In the procedure proposed by \cite{PintoPereira}, the initial term in the linear predictor is the constant term, but in the case of experiments involving mixtures, in most cases, this may not be true. To verify the initialization term in the linear predictor, we must perform a hypothesis test to know whether the constant term exists or not in the linear predictor; see \cite{MarquardtSnee}. Thus, we will need to check if the parameters related to the linear terms (main effects) are equal; If they are equal, it means that the initial model must contain only the constant term; otherwise, the initial model will not have the constant term and must contain all of the main terms of mixture effects. Therefore, if we have $a$ mixture variables given by, $x_{1},\ldots,x_{a}$, the linear predictor containing the main effects will be $\eta=\beta_{1}x_{1} +\ldots+\beta_{a}x_{a}$; if $\beta_{j}=\beta_{0}$ for all $j=1,\ldots,a$, then $\eta=\beta_{0}\left(\sum_{j=1}^{a }x_j\right)=\beta_{0}$. Thus, if we accept that $\beta_{j}=\beta_{0}$ for all $j=1,\ldots,a$, then the initial linear predictor will be $\eta=\beta_{0}$ and  there will not be terms involving the main effects $x_{1},\ldots,x_{a}$. If we reject that $\beta_{j}=\beta_{0}$ for some $j$, then the initial linear predictor will contain all the main effects, that is, $\eta=\beta_{1}x_{1}+ \ldots+\beta_{a}x_{a}$, see \cite{MarquardtSnee} for details.

The hypothesis test to verify the existence of a constant term in the linear predictor will be carried out as follows. We want to test whether the parameters related to the main effects, $x_{1}, \ldots,x_{a}$ are equal to or different from a constant value, say, $\beta_{0}$; for this, we consider the hypotheses $H_{0}: \beta_{i}=\beta_{0}$ for all $i=1,\ldots,a$, versus $H_{1}: \beta_{i} \neq \beta_{0}$ for some $i$, with $i=1,\ldots,a$. Note that under $H_{0}$, the predictor $\eta_{0}$, considering only the main effects, can be written as $\eta_{0}=\beta_{0}$; under $H_{1}$, the linear predictor is given by $\eta_{1}=\sum_{i=1}^{a}\beta_{i}x_{i}$. Also note that, due to the mixture constraint (\ref{eq1}) and considering, for example, $x_{a}$ as a slack variable (see Section \ref{MixtureExperiments}), the linear predictor $\eta_1$ can be written as $\eta_{1}^{*}=\beta_{a}+(\beta_1-\beta_{a})x_{1}+\ldots+(\beta_{(a-1)}-\beta_{a })x_{(a-1)}$ $=\beta_{0}^{*}+\beta_{1}^{*}x_{1}+\ldots+\beta_{(a-1)}^{ *}x_{(a-1)}$. Thus, $\eta_{0} \subset \eta^{*}_{1}$, where, $\eta_{1}^{*}$ corresponds to the predictor $\eta_{0}$ plus the terms $x_1 ,\ldots,x_{(a-1)}$. Consequently, since we now have two nested models, the hypothesis tests used in Algorithm 3 can be applied. Note that the choice of the slack variable is arbitrary, because it does not affect the functioning of hypothesis test.

It is also worth emphasizing that the proposed variable selection process concerns the Scheff\'{e}'s models. As already mentioned, the introduction of slack variables at the beginning of the variable selection process is just an internal process initialization device to verify the existence or not of the constant term in the model. Thus, according to Algorithm 3, after defining the initial linear predictor $\vartheta$, the variable selection process proceeds normally, considering the terms present in Scheff\'{e}'s models.

Algorithm 3, adapted for the case of experiments with mixtures, shows how the terms of the considered model (mean or dispersion) are selected to find the optimal model.

\vspace{0.3cm}

\noindent {\bf Algorithm 3}: Pseudo algorithm for variable selection in mean (dispersion) model 

In the JMMD, for a given and fixed dispersion (mean) model, choose the terms of the mean (dispersion) model, considering the following sequence of steps.

\begin{enumerate}

\item Let $x_1, \ldots, x_a$ be the terms referring to the mixture components, satisfying the mixture restrictions, according to equation (\ref{eq1}), and consider a possible model for startup with linear predictor $\vartheta_0 = \theta_1 x_1 + \ldots + \theta_a x_a$. Do a hypothesis test to check if $H_0: \theta_i = \theta_0$,  $\forall$ $i = 1, \ldots, a$. If $H_0$ is not rejected, do the initial linear predictor $\vartheta = \theta_0$, otherwise $\vartheta = \vartheta_0$.

\item Let $\cal{V}$ be the set of terms that will be used in the variable selection process and, according to step 1, consider $\vartheta$ the linear predictor of the initial model. Note that the terms in the set $\cal{V}$ will also depend on the test performed in step 1. If $H_0$ has not been rejected, the set $\cal{V}$ cannot contain the terms $x_{1}, \ldots,x_{a}$. Otherwise, $\cal{V}$ must not contain the constant term.

\item Fit models with linear predictors $\vartheta_j=\vartheta +\theta_j v_j$, for all $v_j \in \cal{V}$. 
 
\item For each model fitted in step 3, calculate the appropriate selection measure to check the quality of fit and choose the model with linear predictor, say $\vartheta_k=\vartheta + \theta_k v_k$, that has the best fit. Evaluate the value of the selection measure found in the current iteration with the value of the selection measure found in the previous iteration. If the value of the current selection measure is better than the previous value, go to step 5, otherwise, go to step 6.
  
\item For those models with nested linear predictors $\vartheta$ and $\vartheta_k$, apply an appropriate test to verify the significance of the addition of $v_k$ into the linear predictor $\vartheta$. If $v_k$ is significant, remove the term $v_k$ from $\cal{V}$, do $\vartheta=\vartheta_k$ and return to step 3. Otherwise, the procedure stops and $\vartheta$ is the final model chosen.

\item Evaluate, in the same way as in step 5, the significance of the addition of $v_k$ into the linear predictor $\vartheta$. If $v_k$ is significant, do $\vartheta=\vartheta_k$. Stop the procedure and $\vartheta$ is the final model chosen. 

\end{enumerate} 

In step 5 of Algorithm 3, in the case of the dispersion model, for which a GLM is assumed with Gamma distribution and dispersion parameter equal to 2, the test for comparing two nested models is the usual analysis of the deviance for GLMs. In the case of the mean model, where $\mat{\phi}$ is given, the test for comparing two nested models is the $F$-test. Thus, if $H_c$ and $H_d$ are two nested hypothesis of dimension $c < d$, that is $\eta_c \subset \eta_d$, then, under $H_c$, 

\begin{equation}
\label{eqn:F}  
\frac{\left(\sum_{i=1}^{n}\frac{1}{\phi_i} d^{*}_{\mu_{c_i}} - \sum_{i=1}^{n}\frac{1}{\phi_i} d^{*}_{\mu_{d_i}} \right) \Big/ (d-c)}{\left(\sum_{i=1}^{n}\frac{1}{\phi_i} d^{*}_{\mu_{d_i}} \right) \Big/(n-d)}
\end{equation}

\noindent has an asymptotic $F_{d-c,n-d}$, where $d^{*}_{\mu_{k_i}}$ is the standardized deviance component in relation to $\mu_{k_i}$, given in Table 1. 

The result in equation (\ref{eqn:F}) can be shown in the following way. For a given $\phi_i$, apart from the constant, the extended quasi-likelihood is the quasi-likelihood for a model with variance function $V(\mu_i)$ and with prior weights $\frac{1}{\phi_i}$ \citep{LeeNelderI}. However, in the iterative process to build joint models of the mean and dispersion there is some uncertainty in the estimation of $\phi_i$, thus \cite{PintoPereira} assume that $\phi_i$ be replaced by $\tau \phi_i$, where $\tau$ is an unknown constant. In this way, if $H_c$ and $H_d$ are two nested hypothesis of dimension $c < d$, that is $\eta_c \subset \eta_d$, then, considering the quasi-likelihood ratio statistic and according to \cite{McCullagh}, under $H_c$ the change in the extended quasi-deviance, given by $-2\left\{Q^{+}(\mat{\mu}_c,\mat{\phi};\mat{y})- Q^{+}(\mat{\mu}_d,\mat{\phi};\mat{y}) \right\}= \frac{1}{\tau}\left[ \sum_{i=1}^{n}\frac{1}{\phi_i} d^{*}_{\mu_{c_i}} - \sum_{i=1}^{n}\frac{1}{\phi_i} d^{*}_{\mu_{d_i}}\right]$, has an asymptotic Chi-square distribution with $d - c$ degrees of freedom. On the other hand, when there exists a distribution of the exponential family with a given variance function, it turns out that the extended quasi-likelihood is the saddlepoint approximation to that distribution \citep{NelderLee}. Thus, as pointed out by \cite{DunnSmyth}, p. 277, when the saddlepoint holds, the residual deviance has approximately a Chi-square distribution with $n - n_p$ degrees of freedom, where $n_p$ is the number of parameters of the considered model. In our case, considering the linear predictor $\eta_d$ under the hypothesis $H_d$, the residual deviance is given by $\frac{1}{\tau}\sum_{i=1}^{n}\frac{1}{\phi_i} d^{*}_{\mu_{d_i}}$. Therefore, according to \cite{DunnSmyth}, $\frac{1}{\tau}\sum_{i=1}^{n}\frac{1}{\phi_i} d^{*}_{\mu_{d_i}}$ has a Chi-square distribution with $n-d$ degrees of freedom. In this way, the result in equation (\ref{eqn:F}) is obtained directly by applying the definition of the $F$ distribution. Note that the $F$-test does not depend on the parameter $\tau$. 

The criteria we used to check the quality of the model's fit in the variable selection procedure are the same ones used by \cite{PintoPereira} and are presented below.

For the mean model we used the extended Akaike information criterion ($EAIC$), proposed by \cite{WangZhang}, is given by

\begin{equation}
\label{eq.eaic}
 EAIC=-2Q^{+}(\mat{\mu},\mat{\phi};\mat{y}) + F(\kappa,n),
\end{equation}

\noindent where $\kappa=p+q$ is the sum of the number of parameters of the mean and dispersion models and $F(\kappa,n)$ is a  penalty function that depends on $\kappa$ and $n$. The penalty we used was $F(\kappa,n)=\frac{2 \kappa n}{n-\kappa-1}$.

For the dispersion model we used the corrected Akaike information criterion, given by $AIC_c = -2\ln \hat{L} + \frac{2qn}{n-q-1}$, where $\hat{L}$ is the maximum value of the likelihood function. For both $EAIC$ and $AIC_c$ criteria, the lower the value, the better.

In addition to these criteria, the criteria proposed by \cite{PintoPereira} were also used. The goodness of fit criteria, proposed by them for the mean and dispersion models in the JMMD, were as follows. For the mean model they proposed the criterion given by

\begin{equation} 
  \label{eqn:r2m} 
\footnotesize{  
  \tilde{R}_m^2= 1- \frac{\sum_{i=1}^{n} \tilde{d}_{V_i}(y_i,\widehat{\mu}_i)/(n-\lambda_{n}p) }{\sum_{i=1}^{n} \tilde{d}_{V_i}(y_i,\overline{y})/(n-1)}}
\end{equation}

\noindent and for the dispersion model the proposed criterion was

\begin{equation} 
  \label{eqn:r2d} 
\footnotesize{  
  \tilde{R}_d^2= 1- \frac{\sum_{i=1}^{n} d_{V_i}(d_i^{*},\widehat{\phi}_i)/(n-\lambda_{n}q) }{\sum_{i=1}^{n} d_{V_i}(d_i^{*},\overline{d^{*}})/(n-1)}},
\end{equation}

\noindent where $\tilde{d}_{V}(a,b)=\left[ \int_{a}^{b}\sqrt{1+ \phi^{2}[V'(t)]^2}dt \right]^2$, $d_{V}(a,b)= \left[ \int_{a}^{b}\sqrt{1+[V'(t)]^2}dt \right]^2$, with $V'(t)=\frac{dV(t)}{dt}$ and $V(t)$ is a continuous and derivable function in $(a, b)$, $n$ is the number of observations, $\lambda_n$ is a function of $n$, $p$ is the number of parameters in the mean model, $q$ is the number of parameters in the dispersion model, $\hat{\mu}_i=\varphi^{-1}(\hat{\eta}_i)$, $\hat{\phi}_i=\exp(\hat{\xi}_i)$, $\overline{y}=(1/n)\sum_{i=1}^n y_{i}$ and $\overline{d^{*}}=(1/n)\sum_{i=1}^n d_{i}^{*}$, with $y_i$ and $d_i^{*}$ given in Table \ref{Tab1}.  For both $\tilde{R}_m^2$ and $\tilde{R}_d^2$ criteria, the higher the value, the better.

Note that the variable selection procedure conducted by Algorithm 2 also provides a way to check whether the dispersion is constant or not. That is, if the final dispersion model does not contain explanatory variables, then the model will only have the constant term and, therefore, the dispersion is constant; otherwise we have non-constant dispersion.

\subsection{Mean and variance models} 
\label{Mean_Variance_Models}

After obtaining the final joint model for mean and dispersion we can get the models for $E(Y)$ and $Var(Y)$; mean and variance, respectively. From Table \ref{Tab1} we can see that the mean model is directly obtained by $E(Y)=\varphi^{-1}(\eta)$, while the model for variance is given by $Var(Y)=\exp({\xi}) V\left[\varphi^{-1}(\eta)\right]$. 

For the case where the mixture experiment involves noise variables, the results of the experiment are conditioned on the noise variables, which are considered as random variables. In this case, the mean and variance models, previously found by the method of joint modeling of mean and dispersion, are now conditioned to the noise variables and given by $E(Y|\mat{N})$ and $Var(Y|\mat{N})$, where $\mat{N}$ is a vector of random noise variables. We can get the unconditional models for $E(Y)$ and $Var(Y)$ using the expressions $E(Y)=E(E(Y|\mat{N})) $ and $Var(Y)= Var(E(Y|\mat{N}))+ E(Var(Y|\mat{N}))$. It is assumed that the probability distribution or at least the mean and variance of $\mat{N}$ are known or can be estimated. 

It is worth mentioning that, in a problem of robust design, see \cite{Taguchi}, knowing the models of mean and variance, we can minimize the variability by finding the optimum settings for the explanatory variables that affect the variance model and then adjusting the mean to its target value by finding appropriate settings for the explanatory variables that affect the mean model. For the case of mixture experiments, this process leads to an optimization problem, in which the mixture constraint must be considered, in addition to other restrictions on the mixture variables, if  any.

\section{Assessment and validation of the variable selection procedure}
\label{NumericalEvaluation}

To evaluate the accuracy of the variable selection procedure in mixture experiments, Monte Carlo simulations were performed in order to verify the percentage of hits. 

In the evaluation process of the selected models, three possible category models were considered:

\begin{itemize}
    \item[-] Category 1 (Type 1 model) is defined as the selected models that contain the same parameters of the simulated model plus other terms. This category is associated to type I error in statistical hypothesis testing of parameters, representing the occurrence of this error in at least one of the tests described in step 5 of Algorithm 3. This is one of the categories that we consider acceptable because they contain all the terms of the simulated model. Even though our objective is to obtain correct models, we cannot neglect the percentage of models in this category, as a way of expressing the quality of the procedure. In some applications, correct models may have hard to obtain and models in this category can be useful. 
  
  \item[-] Category 2 (Type 2 model) is defined as the selected models that not contains all the terms of the simulated model. This category is associated to type II error in statistical hypothesis testing of parameters, representing the occurrence of this error in at least one of the tests described in step 5 of Algorithm 3. Models in this category will be considered as poor models, not being included in the list of acceptable models. It is extremely important that the criteria adopted have a low risk of obtaining models of this type.
  
  \item[-] Correct models is defined as the selected models that contains exactly the same parameters of the simulated model. This category is associated to no error on the variable selection procedure. 
  
  \item[-] Acceptable models is defined as the selected models that contains exactly the same parameters of the simulated model (Correct models) or plus other terms (Type 1 models). Therefore, it represents the union of two categories (Type 1 or Correct). In our study, although presenting the results separately for the Correct and Type 1 models, we also chose to present the results joining these two categories, which can generate a proportion of useful models in several applications in which the costs associated with the factors are not so relevant. 
  
\end{itemize}

 All simulations were performed using the statistical software R \citep{RTeam}. The simulations were performed for mean models in which the response variable in the $i$th trial, $Y_i$, was considered to follow the Normal distribution, i.e., with $V(\mu_i)=1$. In order to ensure non-constant dispersion, the data simulation process must ensure that $E(Y_i)=\mu_i$ and $Var(Y_i)=\phi_i$.  Such results can be directly obtained by taking $Y_i\sim N(\mu_i,\phi_i)$, where $\mu_i=\eta_i$ and $\phi_i=\exp\{\xi_i\}$, with $\eta_i$ and $\xi_i$ given, respectively, by the equations (\ref{eqn:m1}) and (\ref{eqn:d1}). For the dispersion model, the response variable, $d^{*}_i$, was considered to have a Gamma distribution, i.e., $V(\phi_i)=\phi_{i}^{2}$, and with logarithm link function, according to Table \ref{Tab1}. The linear predictors, chosen for the data generation process, were:

\begin{equation}
  \label{eqn:m1}  
  \eta_i= \beta_1x_{1i}+\beta_2x_{2i}+\beta_3x_{3i} + \beta_{13}x_{1i} x_{3i} + \gamma_{3z}x_{3i} z_{i}
\end{equation} and

\begin{equation}
 \label{eqn:d1}   
  \xi_i= \gamma_1x_{1i}+\gamma_2x_{2i}+\gamma_{13}x_{1i}x_{3i}, 
\end{equation}

\noindent where the linear predictor shown in equation (\ref{eqn:m1}) was used for the mean model and the linear predictor shown in equation (\ref{eqn:d1}) was used for the dispersion model. 
For the simulation process, a factorial experiment was considered in the process variable $z$, crossed with a simplex-centroid design in the mixture variables $x_{1}$, $x_{2}$, $x_{3}$. The values of $(x_{1},x_{2},x_{3})$,  obtained from  three component simplex-centroid design, consisted of the points (1,0,0), (0,1,0), (0,0,1), (1/2, 1,2, 0), (1/2, 0, 1/2), (0, 1/2, 1/2), (1/3, 1/3, 1/3). The $z$ values were fixed at $-1$ and $1$. Thus, the experiment consisted of 14 experimental runs, where for each of the values of $z= -1$ and $z=1$ the seven mixture combinations of the simplex-centroid design were performed. The numbers of replications of the experiment considered in the simulation process were 2, 4, 7 and 11, totaling the following sample sizes 28, 56, 98 and 154. The replication numbers were chosen in order to have small, medium and large sample sizes.

The values chosen for the parameter vector $(\beta_1,\beta_2,$ $\beta_3, \beta_{13}, \gamma_{3z}, \gamma_1,\gamma_2,\gamma_3, \gamma_{13} )$, used to perform simulations, were $(15,15,10,35,25,0.4,0.5,0.5,3.5)$. For the simulations, 64 different scenarios were considered, which originated from the combination of sixteen possibilities for the goodness of fit criteria for the mean and dispersion models with four different sample sizes, as shown in Table \ref{table:Freq}. For each scenario considered 1000 Monte Carlo replications were performed. 

Regarding the level of significance, considered for the hypothesis tests in step 5 of Algorithm 3, we understand that a level of $0.05$ is very strict and could exclude important variables from the model. \cite{LeeKoval} examined the question of the level of significance in stepwise procedures for the case of logistic regression and suggested that the level of significance adopted should be between 0.15 and 0.20. Based on the discussions presented by \cite{LeeKoval}, in the references considered there and following \cite{PintoPereira}, we decided to use an intermediate value between $0.05$ and $0.15$, that is, a significance level of $0.10$. In all simulated studies performed, $i = 1,\ldots, n$ refers to the $i$th response observed and $n$ is the sample size. 

To facilitate interpretation, a cluster analysis was performed in order to discriminate the criteria based on the percentage of Type 1, Type 2 and Correct models, creating groups of criteria with similar behaviors. The aim is grouping a set of criteria in such a way that criteria in the same group (cluster) are more similar (in percentage) to each other than to those in other groups. The Euclidean distance was used as the similarity measure, see \cite{JohnsonWichern} for more details. The simulation studies are shown in Table \ref{table:Freq}. The result of the cluster analysis is shown in the last column of the Table \ref{table:Freq}. The analysis suggested the creation of three clusters, called A (three criteria), B (ten criteria)  and C (three criteria), showed in this order.

\setlength\tabcolsep{1.8pt}

\begin{table}[!ht]
\centering
\caption{Percentages of model types found in 1000 Monte Carlo simulations of the variable selection procedure  in mixture experiments}
\scalefont{0.90}
\label{table:Freq}
\begin{tabular}{llclccclclccclclccclc} 
\toprule
\multicolumn{2}{c}{\multirow{2}{*}{Criteria}}           & \multicolumn{18}{c}{Category}                                                                                       & \multirow{3}{*}{Cluster}  \\ 
\cline{3-20}
\multicolumn{2}{c}{}                                    & \multicolumn{6}{c}{Type 1}          & \multicolumn{6}{c}{Correct}            & \multicolumn{6}{c}{Type 2}          &                           \\ 
\cline{1-2}\cline{4-7}\cline{10-13}\cline{16-19}
Mean                       & Dispersion                 &  & n=28  & n=56  & n=98 & n=154 &  &  & n=28  & n=56  & n=98 & n=154 &  &  & n=28  & n=56  & n=98 & n=154 &  &                           \\ 
\hline
$\tilde{R}_m^2(1)$       & $AIC_c$                                                      &  & 0.427 & 0.528 & 0.504 & 0.501 &  &  & 0.317 & 0.398 & 0.410 & 0.405 &  &  & 0.256 & 0.074 & 0.086 & 0.094 &  & \multirow{3}{*}{A}        \\
$\tilde{R}_m^2(\sqrt n)$ & $AIC_c$                                                      &  & 0.177 & 0.547 & 0.579 & 0.525 &  &  & 0.384 & 0.377 & 0.363 & 0.385 &  &  & 0.439 & 0.076 & 0.058 & 0.090  &  &                           \\
$\tilde{R}_m^2(\log n)$  & $AIC_c$                                                      &  & 0.412 & 0.543 & 0.503 & 0.479 &  &  & 0.284 & 0.389 & 0.412 & 0.431 &  &  & 0.304 & 0.068 & 0.085 & 0.090  &  &                           \\ 
\hline
$EAIC$                   & $AIC_c$                                                      &  & 0.262 & 0.459 & 0.433 & 0.447 &  &  & 0.377 & 0.467 & 0.474 & 0.461 &  &  & 0.361 & 0.074 & 0.093 & 0.092 &  & \multirow{10}{*}{B}       \\
$\tilde{R}_m^2(1)$       & $\tilde{R}_d^2(1)$                                           &  & 0.331 & 0.425 & 0.398 & 0.422 &  &  & 0.414 & 0.491 & 0.528 & 0.513 &  &  & 0.255 & 0.084 & 0.074 & 0.065 &  &                           \\
$\tilde{R}_m^2(\log n)$  & $\tilde{R}_d^2(1)$                                           &  & 0.062 & 0.325 & 0.367 & 0.383 &  &  & 0.487 & 0.549 & 0.556 & 0.539 &  &  & 0.451 & 0.126 & 0.077 & 0.078 &  &                           \\
$EAIC$                   & $\tilde{R}_d^2(1)$                                           &  & 0.331 & 0.432 & 0.424 & 0.408 &  &  & 0.373 & 0.478 & 0.510 & 0.521 &  &  & 0.296 & 0.090 & 0.066 & 0.071 &  &                           \\
$\tilde{R}_m^2(1)$       & $\tilde{R}_d^2(\sqrt n)$                                     &  & 0.190  & 0.359 & 0.409 & 0.415 &  &  & 0.366 & 0.514 & 0.510 & 0.496 &  &  & 0.444 & 0.127 & 0.081 & 0.089 &  &                           \\
$\tilde{R}_m^2(\log n)$  & $\tilde{R}_d^2(\sqrt n)$                                     &  & 0.333 & 0.425 & 0.489 & 0.454 &  &  & 0.271 & 0.465 & 0.445 & 0.485 &  &  & 0.396 & 0.110 & 0.066 & 0.061 &  &                           \\
$EAIC$                   & $\tilde{R}_d^2(\sqrt n)$                                     &  & 0.066 & 0.200   & 0.328 & 0.407 &  &  & 0.288 & 0.534 & 0.572 & 0.530 &  &  & 0.646 & 0.266 & 0.100 & 0.063 &  &                           \\
$\tilde{R}_m^2(1)$       & $\tilde{R}_d^2(\log n)$                                      &  & 0.240  & 0.490  & 0.448 & 0.464 &  &  & 0.318 & 0.408 & 0.494 & 0.486 &  &  & 0.442 & 0.102 & 0.058 & 0.050 &  &                           \\
$\tilde{R}_m^2(\log n)$  & $\tilde{R}_d^2(\log n)$                                      &  & 0.199 & 0.451 & 0.452 & 0.453 &  &  & 0.227 & 0.417 & 0.473 & 0.473 &  &  & 0.574 & 0.132 & 0.075 & 0.074 &  &                           \\
$EAIC$                   & $\tilde{R}_d^2(\log n)$                                      &  & 0.300   & 0.414 & 0.436 & 0.442 &  &  & 0.348 & 0.492 & 0.495 & 0.495 &  &  & 0.352 & 0.094 & 0.069 & 0.063 &  &                           \\ 
\hline
$\tilde{R}_m^2(\sqrt n)$ & $\tilde{R}_d^2(1)$                                           &  & 0.036 & 0.217 & 0.343 & 0.394 &  &  & 0.218 & 0.584 & 0.573 & 0.532 &  &  & 0.746 & 0.199 & 0.084 & 0.074 &  & \multirow{3}{*}{C}        \\
$\tilde{R}_m^2(\sqrt n)$ & $\tilde{R}_d^2(\sqrt n)$                                     &  & 0.266 & 0.406 & 0.446 & 0.424 &  &  & 0.320 & 0.477 & 0.485 & 0.511 &  &  & 0.414 & 0.117 & 0.069 & 0.065 &  &                           \\
$\tilde{R}_m^2(\sqrt n)$ & $\tilde{R}_d^2(\log n)$                                      &  & 0.176 & 0.411 & 0.441 & 0.441 &  &  & 0.283 & 0.445 & 0.479 & 0.479 &  &  & 0.541 & 0.144 & 0.080  & 0.080 &  &                           \\
\bottomrule
\end{tabular}
\end{table}

\begin{figure}[!ht]
\centering
\includegraphics[scale=0.50]{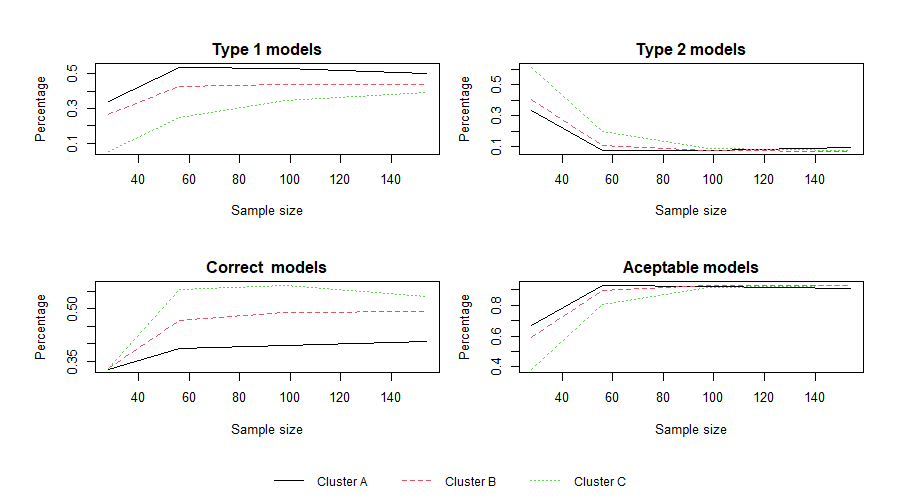}
\caption{Graphs showing the clusters mean percentage of Type 1, Type 2, Correct and Acceptable models found in 1000 Monte Carlo simulations of the variable selection procedure in mixture experiments. The sample size considered were 28, 56, 98 and 154.}
\label{fig:graphics-sim}
\end{figure}

Graphics by clusters for Type 1, Type 2, Correct and Acceptable models are shown in Figure \ref{fig:graphics-sim}, where the information for each cluster in each sample size was reduced to the arithmetic mean of the cluster components. The graph called Acceptable models shows the percentages of Correct models added to Type 1 models.

The results of Table \ref{table:Freq} and Figure \ref{fig:graphics-sim} clearly suggest that the percentage of acceptable models is associated with the sample size; the larger, the better the percentage of acceptable models. From sample size equal to 56, the percentages of acceptable models reach high levels, with small differences between the criteria, indicating that the procedure provides great results even for intermediate samples, regardless of the criteria adopted. These results suggest that, for acceptable models, the adopted criterion is not an important factor when we have large samples, since the procedure is robust to find acceptable models with high probability. 

However, the same does not happen when we consider only the correct models. We noticed that, for sample sizes above 56, we have better results when we use the criteria associated with cluster C. The criteria included in this cluster have something in common, i.e., all of them use $\tilde{R}_{m}^2(\sqrt n)$ to testing parameters associated with the mean model and does not use the $AIC_c$ to testing parameters associated with the dispersion model. This may be an indication that, for large samples and considering only correct models as a target, the use of $\tilde{R}_{m}^2(\sqrt n)$ for mean models avoiding the $AIC_c$ for dispersion models should be encouraged.  

For sample size equal to 28, we have a different scenario. We noticed that there is no cluster capable of increasing the proportion of correct models, but there is a cluster capable of increasing the proportion of Type 1 models and consequently increasing the proportion of acceptable models. This cluster is A; which also decreases the proportion of Type 2 models. The criteria included in cluster A always use $AIC_c$ to testing parameters associated with the dispersion model and  does not use the $EAIC$ to testing parameters associated with the mean model. On this way, for small samples this may indicate that the use of $AIC_c$ for dispersion models avoiding the $EAIC$ for mean models would be the best option.

\section{Application to a bread-making problem}
\label{Application}

The bread-making problem, originally presented by \cite{FaergestadNaes}, according to \cite{NaesFaergestadCornell}, consisted of an experiment with three ingredients of mixture and two noise variables, and had as objective to investigate and to value the final quality of flour, composed by different mixtures of wheat flour, for production of bread. Also, according to \cite{NaesFaergestadCornell}, \cite{FaergestadNaes} considered three types of wheat flour: two Norwegian, Tjalve $(x_{1})$ and Folke $(x_{2})$ and one American, Hard Red Spring $(x_{3})$, that were considered as control variables, and two types of process variables: mixing time $(z_{1})$ and the proofing (resting) time of the dough $(z_{2})$, considered as noise variables. The response variable was considered as the loaf volume after baking with target value of 530 ml. The flour blends were considered to be mixing ingredients with $x_{1} + x_{2} +x_{3} = 1$ and with constraints $0.25\le x_{1} \le 1.0$; $0 \le x_{2} \le 0.75$ and $0 \le x_{3} \le 0.75$, where $x_{1}$, $x_{2}$ and $x_{3}$ are the proportions of Tjalve, Folke and Hard Red Spring flour, respectively. For the noise variables, it was considered three situations for the mixing time: 5, 15 and 25 minutes and also three situations for proofing time: 35, 47.5, and 60 minutes. The noise variables were coded as $-1$, 0, 1 according to their minimum, mean and maximum values. 

\begin{figure}[!htb]
\centering
\includegraphics[scale=0.75]{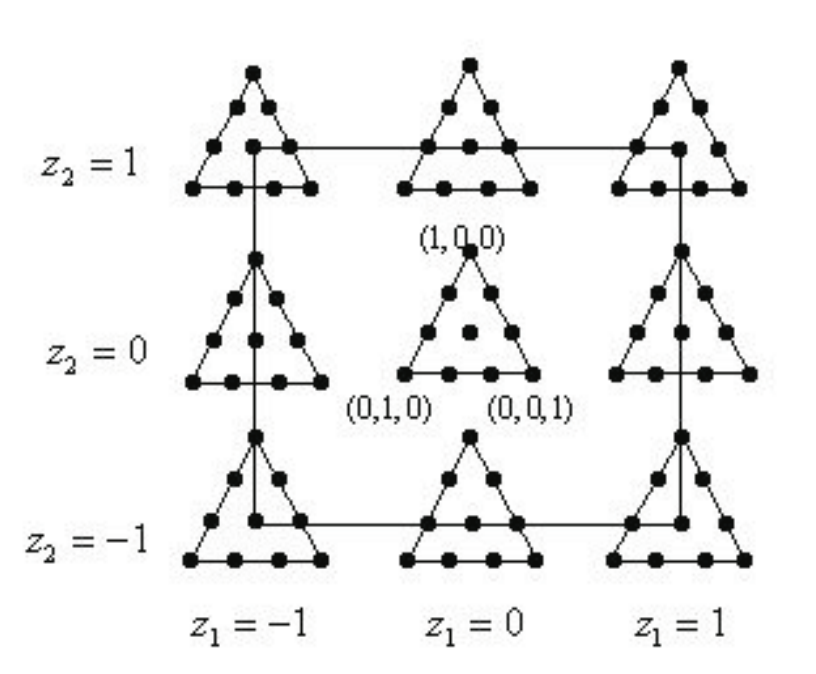}
\caption{Full factorial $3^2$ in two process variables crossed with a third order simplex-centroid design in three mixture components}
\label{Fact_Simplex}
\end{figure}

A full $3^{2}$ factorial design was used for the noise variables and the 10 runs corresponding to a simplex-lattice design were replicated at each of the nine combinations of the mixing and proofing times, so the complete design involved 90 experimental runs as shown in Figure \ref{Fact_Simplex}. 

The volumes recorded for the 10 flour types and the 9 combinations of the noise variables are reproduced in Table \ref{LoafVol}. Additional details of the way in which the experiment was conducted are given by \cite{NaesFaergestadCornell}, and further description of the practical aspects of the study is provided by \cite{FaergestadNaes} according to \cite{NaesFaergestadCornell}.

\begin {table}[!htb]
\caption{Loaf volume for the 10 flour types and the 9 combinations of the noise variables}
\scalefont{0.80}
\label{LoafVol}
\centering
\begin{tabular}{ccccccccccccccc}
\hline                          
& & & & & &\multicolumn{9}{c}{Noise factors} \\
\cline{7-15}
&\multicolumn{3}{c}{Design factors}& &$z_{1}$ & -1 &  0 &  1 &-1 & 0 & 1 & -1& 0 & 1 \\
\cline{2-4}  
$\textrm{n}^{\textrm{o}}$& $x_{1}$& $x_{2}$& $x_{3}$& & $z_{2}$ & -1 & -1 & -1 & 0 & 0 & 0 & 1 & 1 & 1 \\
\hline
1 & 0.25 & 0.75 & 0.00 & & & 378.89 & 396.67 &392.22&445.56&452.22&487.78&457.22&500.56&472.78\\  
2 & 0.50 & 0.50 & 0.00 & & & 388.89 & 423.33 &416.11&460.00&488.89&475.78&472.78&478.00&506.11\\  
3 & 0.75 & 0.25 & 0.00 & & & 426.11 & 483.33 &389.44&474.44&514.44&462.78&506.67&591.67&522.22\\  
4 & 1.00 & 0.00 & 0.00 & & & 386.11 & 459.11 &423.33&458.33&506.11&514.44&545.56&522.22&551.11\\  
5 & 0.25 & 0.50 & 0.25 & & & 417.78 & 437.22 &444.56&484.44&490.00&495.00&497.78&531.11&577.78\\  
6 & 0.50 & 0.25 & 0.25 & & & 389.44 & 447.22 &415.00&490.89&528.89&507.78&517.78&567.22&538.33\\  
7 & 0.75 & 0.00 & 0.25 & & & 448.33 & 459.44 &455.56&436.00&535.00&552.22&507.44&578.89&590.00\\  
8 & 0.25 & 0.25 & 0.50 & & & 413.89 & 485.56 &462.22&483.89&529.44&540.00&565.00&598.89&580.56\\  
9 & 0.50 & 0.00 & 0.50 & & & 415.56 & 514.44 &437.78&493.89&583.33&578.89&524.44&694.44&640.00\\  
10& 0.25 & 0.00 & 0.75 & & & 432.78 & 498.33 &517.22&474.44&568.33&579.44&541.11&638.89&638.89\\  
\hline
\end{tabular}
\end{table}

\cite{NaesFaergestadCornell}, considering Normal distribution and constant variance, carried out a study of the use of robust design methodology to the bread-making problem to investigate the underlying relationships between the response variable loaf volume and the mixture and noise variables, comparing three techniques for analysing the loaf volume, i.e., the mean square error, the analysis of variance and the regression approach, where all factors, the three mixtures components and the two noise variables, were modeled simultaneously. In the analysis considered by \cite{NaesFaergestadCornell} the dispersion was considered constant. However, when we use our variable selection procedure for this data set, we find that the dispersion is not constant. Thus, in this section, we reexamine the bread-making problem considering the possibility of obtaining, in addition to the mean model, a model for variance. For this, we will use the joint modeling of mean and dispersion, presented in Section \ref{JMMD}.

For the application of the variable selection process, we considered $V(\mu) = 1$ and identity link function for the mean model. For the dispersion model we considered the Gamma distribution with a logarithmic link function. The terms of the set $\cal{V}$, used in both the mean and the dispersion models, consisted of the terms of the cubic model in the mixture components crossed with the quadratic model in the noise variables.

In our application, the criterion considered for the mean model was the $\tilde R_m^2$ with $\lambda_n = \sqrt{n}$ and for the dispersion model the criterion was the $\tilde R_d^2$ with $\lambda_n = 1$. These criteria were chosen due to the fact that we have a large sample, which indicates that the criteria in cluster C would be better. Among the criteria of cluster C, we chose those that have the highest percentage of correct models for large samples, according to Table \ref{table:Freq}. 

Table \ref{Steps_Procedure} presents the results of all steps of the variable selection process (Algorithms 2 and 3) taken to obtain the final joint model for the mean and dispersion. The initial test, presented at the beginning of each iteration, for both the mean model and the dispersion model, concerns the hypothesis test to verify whether the model should have a constant term or not; thus, the null hypothesis states that the parameters of the main effects terms are equal to a constant value. In each iteration of Algorithm 2, Algorithm 3 was used to find the terms of the mean and dispersion models. 

\begin{table}[!htp]
\scalefont{0.85}
\centering
\caption{Steps of variable selection algorithm for the JMMD applied to data from the bread-making problem.}
\label{Steps_Procedure}
\begin{tabular}{llccccc} 
\hline  \hline
\multicolumn{1}{l|}{Iteration$^{\dag}$}         & Dispersion model & $\tilde{R}_{d}^2 (1)$                        & $D$ $^\ddag$          & $\chi^2_c$ value$^\star$ & $Pr(>\chi^2_c)$ &            \\ \hline
\multicolumn{1}{l|}{\multirow{8}{*}{1}}  & 1                &  -                            &  -                   &  -                        &  -               &            \\ \cline{2-7} 
\multicolumn{1}{l|}{}                    & Mean model       & $\tilde{R}_{m}^2 (\sqrt{n})$ & $\tilde{R}_{m}^2 (1)$ & $D^{*}$ $^\ddag$         & $F_c$ value$^{\diamond}$     & $Pr(>F_c)$ \\ \cline{2-7} 
\multicolumn{1}{l|}{}       & Initial test                & -                       & -                & -                  & -          & 0.0000     \\
\multicolumn{1}{l|}{}       & $x_1 + x_2 + x_3$             & 0.9810             & 0.9865                & 305938.54                 & -           & -              \\
\multicolumn{1}{l|}{}       & $x_1 + x_2 + x_3+x_1z_2$      & 0.9893             & 0.9935                & 148184.67                  & 91.55         & 0.0000     \\
\multicolumn{1}{l|}{}     & $x_1+x_2+x3+x_1z_2+x_3z_2$    & 0.9901             & 0.9951                & 113114.00                  & 26.35         & 0.0000     \\
\multicolumn{1}{l|}{}    & $x_1+x2+x_3+x_1z_2+x_3z_2+x_1x_3z_1$      & 0.9911                      & 0.9965               & 80267.00                   & 34.37         & 0.0000     \\
\multicolumn{1}{l|}{}    & $\mat{x_1+x_2+x_3+x_1z_2+x_3z_2+x_1x_3z_1+x_2z_2}$    & 0.9888      & 0.9968                & 72773.67         & 8.550          & 0.0045     \\  \hline \hline

\multicolumn{1}{l|}{Iteration$^{\dag}$}         & Dispersion model &  $\tilde{R}_{d}^2 (1)$                        & $D$ $^\ddag$          & $\chi^2_c$ value$^\star$ & $Pr(>\chi^2_c)$ &            \\ \hline
\multicolumn{1}{l|}{\multirow{12}{*}{2}} & Initial test                & -                       & -               & -                   & 0.0330          &            \\
\multicolumn{1}{l|}{}                    & $x_1+x_2+x_3$              & 0.0148                    & 268.68               & -                   & -          &            \\
\multicolumn{1}{l|}{}                    & $\mat{x_1+x_2+x_3+x_2x_3}$        & 0.0319                  & 259.14         & 4.77        &  0.03         &            \\
\multicolumn{1}{l|}{}                    & $x_1+x_2+x_3+x_2x_3+x_1x_3$      & 0.0484                   & 255.76               & 1.69         & 0.19          &             \\ \cline{2-7} 
\multicolumn{1}{l|}{}                    & Mean model       & $\tilde{R}_{m}^2 (\sqrt{n})$ & $\tilde{R}_{m}^2 (1)$ & $D^{*}$ $^\ddag$         & $F_c$ test$^{\diamond}$     & $Pr(>F_c)$ \\ \cline{2-7} 
\multicolumn{1}{l|}{}                    & Initial test                & -                       & -               & -                & -           & 0.0000    \\
\multicolumn{1}{l|}{}                    & $x_1+x_2+x_3$              & 0.9831                        & 0.9880                  & 436.08                  & -          & -     \\
\multicolumn{1}{l|}{}                   & $x_1+x_2+x_3+x_1z_2$     & 0.9911                        & 0.9946                 & 197.54                    & 103.85          & 0.0000     \\
\multicolumn{1}{l|}{}                  & $x_1+x_2+x_3+x_1z_2+x_3z_2$ & 0.9923                        & 0.9962               & 139.80                  & 35.11         & 0.0000     \\
\multicolumn{1}{l|}{}                 & $x_1+x_2+x_3+x_1z_2+x_3z_2+x_1x_3z_1$    & 0.9927                        & 0.9971                & 104.44                  & 28.44          & 0.0000     \\ 
\multicolumn{1}{l|}{}         & $\mat{x_1+x_2+x_3+x_1z_2+x_3z_2+x_1x_3z_1+x_2z_2}$    & 0.9913                     & 0.9975                 & 90.160                   & 13.15           & 0.0005     \\ \hline \hline

\multicolumn{1}{l|} {Iteration$^{\dag}$}                              & Dispersion model &  $\tilde{R}_{d}^2 (1)$                      & $D$ $^\ddag$          & $\chi^2_c$ value$^\star$ & $Pr(>\chi^2_c)$ &            \\ \hline
\multicolumn{1}{l|}{\multirow{10}{*}{3}} & Initial test                & -                       & -               & -                  & 0.9946          &            \\
\multicolumn{1}{l|}{}                    & \bf{1}        & 0.0000       & 268.80               & -                   & -    &            \\
\multicolumn{1}{l|}{}                    & $1+x_1x_3$        & 0.0083                      & 268.36               & 0.22          & 0.64          &            \\ \cline{2-7} 
\multicolumn{1}{l|}{}                    & Mean model       & $\tilde{R}_{m}^2 (\sqrt{n})$ & $\tilde{R}_{m}^2 (1)$ & $D^{*}$ $^\ddag$         & Valor $F_c$ $^{\diamond}$     & $Pr(>F_c)$ \\ \cline{2-7} 
\multicolumn{1}{l|}{}                    & Initial test                & -                       & -                & -                  & -          & 0.0000     \\
\multicolumn{1}{l|}{}           & $x_1+x_2+x_3$      & 0.9810                   & 0.9865                & 305402.98                  & -         & -     \\
\multicolumn{1}{l|}{}           & $x_1+x_2+x_3+x_1z_2$         & 0.9893                       & 0.9935               & 147925.26                  & 91.55         & 0.0000     \\
\multicolumn{1}{l|}{}           & $x_1+x_2+x_3+x_1z_2+x_3z_2$     & 0.9901                       & 0.9951            & 112915.99                   & 26.35         & 0.0000     \\
\multicolumn{1}{l|}{}           & $x_1+x_2+x_3+x_1z_2+x_3z_2+x_1x_3z_1$      & 0.9911                       & 0.9965                & 80126.49              & 34.37        & 0.0000     \\ 
\multicolumn{1}{l|}{}           & $\mat{x_1+x_2+x_3+x_1z_2+x_3z_2+x_1x_3z_1+x_2z_2}$    & 0.9888                     & 0.9968                & 72646.28                   & 8.55          & 0.0045     \\ \hline

\multicolumn{7}{l}{$^{\dag}$ \tiny{In each iteration, the final models for both mean and dispersion are shown in bold.}}\\
\multicolumn{7}{l}{$^{\ddag}$\tiny{$D$ is the deviance for the Gamma model and $D^{*}=\sum_{i=1}^{n}\frac{1}{\phi_i}d_{i}^{*}$.}} \\ 
\multicolumn{7}{l}{$^{\star}$\tiny{$\chi^2_c$ is the value of the Chi-square statistic, calculated from the difference in deviations for two nested Gamma models.}} \\
\multicolumn{7}{l}{$^{\diamond}$\tiny{$F_c$ is the value of the \emph{F}-test statistic, calculated by the equation (\ref{eqn:F}).}} 

\end{tabular}
\end{table}

The additional terms to the current models, which appear sequentially in Table \ref{Steps_Procedure}, are those that, for all terms not belonging to the current model, have the highest value of $\tilde R_m^2$, in the case of the mean model, and the highest value of $\tilde R_d^2$, in the case of the dispersion model. For these terms, candidates to enter the current model, the Chi-square hypothesis tests were used, in the case of the dispersion model, and the $F$ test, in the case of the mean model, to verify if, in fact, the chosen term must be added to the current model.

Measures $D$ and $D^{*}$ were used as test statistics. The values of $D$, $D^{*}$, $\chi^2$, $F$ and $\tilde R_m^2$, with $\lambda_n = \sqrt{n}$ and with $\lambda_n = 1$, are shown for each step of Algorithm 3, where $D$ is the deviance for the Gamma model, $D^{*} = \sum_{i = 1}^{n} \frac{1}{\phi_i}d_{i}^{*}$ (see Algorithm 3), $\chi^2$, the test statistic for the Chi-square test in the dispersion model, is the deviance difference for nested Gamma models and $F$ is the test statistic, given by the equation (\ref{eqn:F}), for the $F$ test in the mean model. It is noteworthy that the measure $\tilde R_m^2$ with $\lambda_n = 1$, which for the normal model is the adjusted $R^2$ of the classical regression, was not used as a selection measure; it has been presented in Table \ref{Steps_Procedure} just to show the impact on the value of $\tilde R_m ^2$, when the number of terms in the model increases, considering the penalty with $\lambda_n = \sqrt{n}$ and with $\lambda_n = 1$. In each iteration, the final models for both mean and dispersion are shown in bold.

Following Algorithm 2, since the value of $\tilde R_m^2$ for the mean model, obtained in the third iteration, was lower than that obtained in the second iteration, the procedure stopped. In this way, three iterations  of Algorithm 2 were performed, obtaining at the end, in the second iteration, a model for the mean with $\tilde R_m^2 (\sqrt{n})$ equal to 0.9913 and $\tilde R_m^2(1)$ equal to 0.9975.

All mean models found in the three iterations of Algorithm 2 had the same terms (see Table \ref{Steps_Procedure}), differing only in the value of the estimates of the effects. The Table \ref{Tab_Examp_Mean_Disp} presents the final estimates, with their standard deviations and Wald tests. Note that all estimates were significant at the 1\% level, except for the term $x_{2}x_{3}$ in the dispersion model, which was significant at the 5\% level.

\begin{table} [!htp]
\scalefont{0.90}
\begin{center}
\caption{Regression coefficients and Wald test for the mean and dispersion models obtained from the bread-making problem data.} 
\begin{tabular}{llllll} \hline
\multicolumn{5}{l}{Mean Model}  \\ \hline \hline

Terms   \hspace{0.3cm}   &  Estimate  \hspace{0.3cm}  & Std. Error \hspace{0.3cm} & $t$  value \hspace{0.3cm}   & $Pr(>|t|)$  \hspace{0.1cm}   \\ \hline 
 $x_{1}$      &  488.961 &  7.263  & 67.323  & 0.0000   \\
 $x_{2}$      &  432.210 &  7.791  & 55.472  & 0.0000   \\
 $x_{3}$      &  574.124 &  9.675  & 59.340  & 0.0000   \\ 
 $x_{1}z_{2}$ &  56.621  &  8.895  & 6.365   & 0.0000   \\
 $x_{3}z_{2}$ &  79.146  &  11.850 & 6.679   & 0.0000   \\
 $x_{2}z_{2}$ &  35.904  &  9.543  & 3.762   & 0.0003   \\
 $x_{1}x_{3}z_{1}$ & 174.216  &  29.706   & 5.865 & 0.0000 \\ \hline
 
 \multicolumn{5}{l}{Dispersion Model}  \\ \hline \hline
 
Terms            &  Estimate    &  Std. Error & $t$ value  & $Pr(>|t|)$      \\ \hline
$x_{1}$         &  6.9984    &  0.3439   & 20.352  & 0.0000 \\
$x_{2}$         &  5.9400    &  0.5607   & 10.594  & 0.0000 \\
$x_{3}$         &  7.3250    &  0.5607   & 13.064  & 0.0000 \\
$x_{2}x_{3}$    & -7.9662    &  3.4523   & -2.307  & 0.0234 \\ \hline 
\end{tabular}
\label{Tab_Examp_Mean_Disp}
\end{center}
\end{table}

\begin{figure}[!ht]
\centering
\includegraphics[scale=0.65]{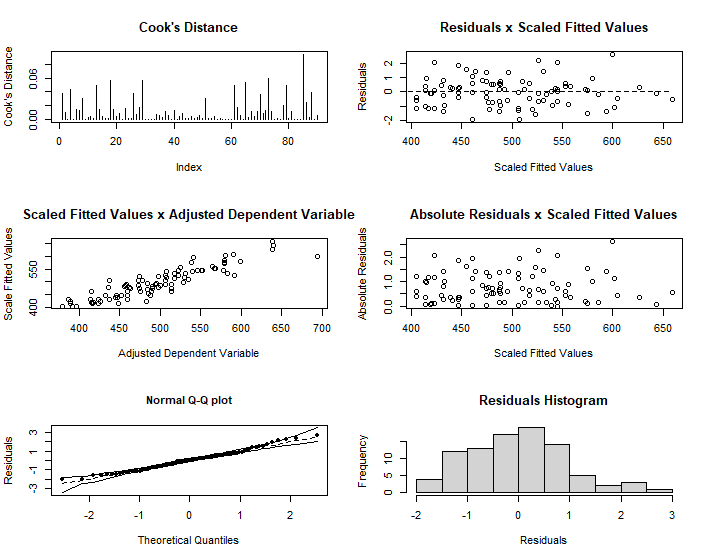}
\caption{Diagnostic plots for the mean model obtained for the bread-making problem. }
\label{Graphics_resid_mean_model}
\end{figure}

The diagnostic plots for the mean model, displayed in Figure \ref{Graphics_resid_mean_model}, show that the model found is adequate. The upper left panel plots Cook's distances. Since all values of Cook's distance do not exceed the value 1, there is no evidence of  outliers. The upper right panel plots the residuals versus the scaled fitted values. The plot indicates that the residuals seem to be distributed around zero with a constant amplitude, indicating the absence of possible anomalies, such as wrong choice of the link function or omission of a quadratic term in some covariate, since there is no curvature present. The middle left panel presents the scaled fitted values versus the adjusted dependent variable. Since the plot appears to show a straight-line pattern, there is no evidence to question the inadequacy of the link function. The middle right panel plots the absolute residuals against the scaled fitted values. Since no trend is found on this plot, there is no evidence of inadequacy of the variance function. The lower left panel displays the Q-Q plot of the residuals, indicating evidence of normality of residuals. This same issue is also verified in the lower right panel that displays the histogram of residuals, which shows that there is no consistent evidence to reject the hypothesis of normality. The type of residual considered in the plots of Figure \ref{Graphics_resid_mean_model} was the deviance residual. For more details on model checking in generalized linear models, see Chapter 12 of  \cite{McCullaghNelder}.

For the dispersion model, the models found in each iteration of Algorithm 2 were different. This is due to the fact that the response  variable $d^*$, is not the same in each iteration of Algorithm 2. The final dispersion model had a value of $\tilde R_d^2(1)$ equal to $0.0319$ (see Table \ref{Steps_Procedure}) and showed significant effects by the Wald test (see Table \ref{Tab_Examp_Mean_Disp}). 

\begin{figure}[!ht]
\centering
\includegraphics[scale=0.5]{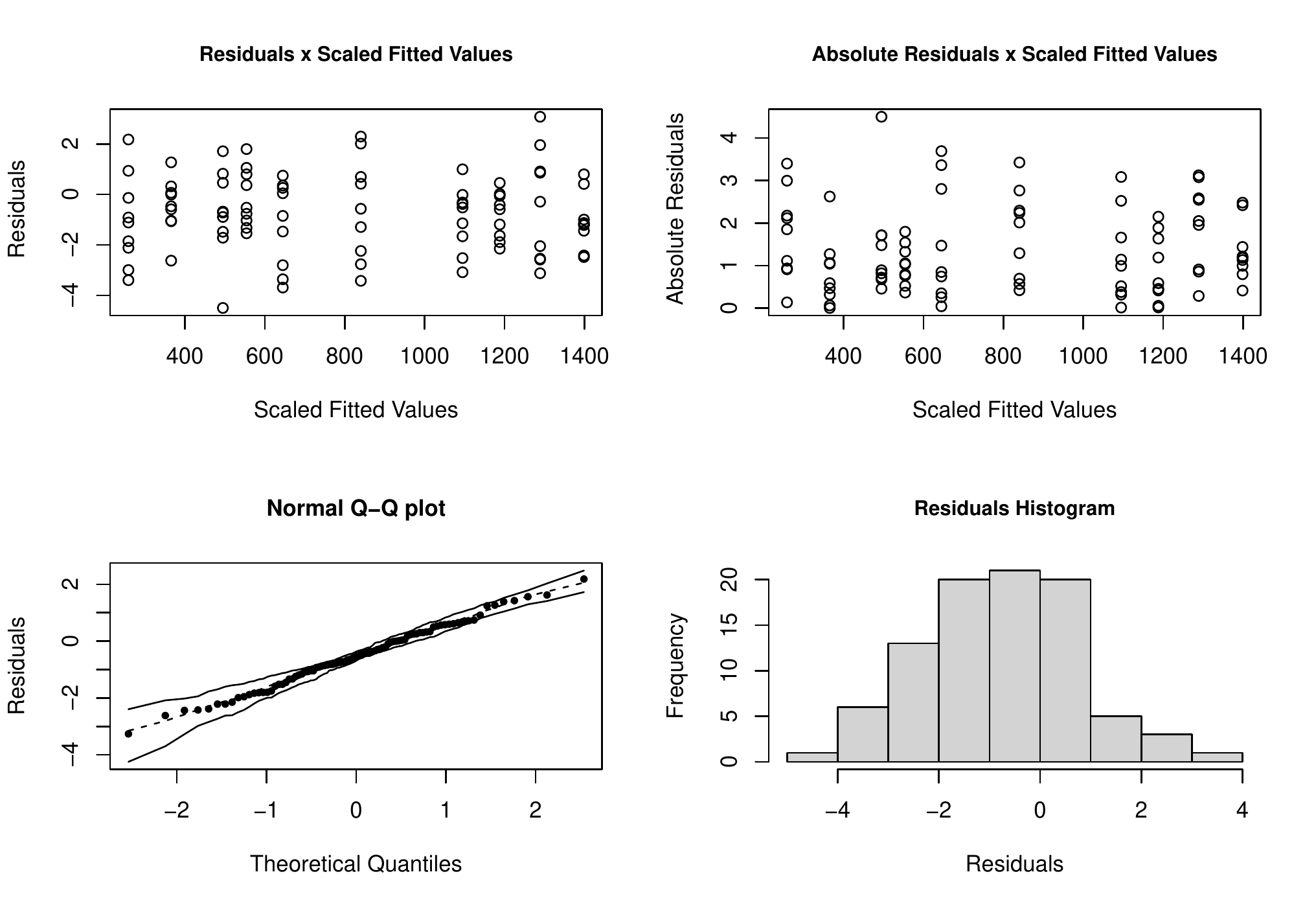}
\caption{Diagnostic plots for the dispersion model obtained for the bread-making problem.}
\label{Graphics_resid_disp_model}
\end{figure}

The diagnostic plots for the dispersion model, displayed in Figure \ref{Graphics_resid_disp_model}, show that the model found is adequate. The upper left panel plots the residuals versus the scaled fitted values. In the same way as analyzed for the mean model, the plot indicates the absence of possible anomalies in the model. The upper right panel plots the absolute residuals against the scaled fitted values. The plot indicates no evidence of inadequacy of the variance function. The lower left panel displays the Q-Q plot of the residuals, indicating evidence of normality of residuals. This same issue is also verified in the lower right panel that displays the histogram of residuals. As commented for the plots of the mean model, the type of residual considered in the plots of Figure \ref{Graphics_resid_disp_model} was the deviance residual. 

The unconditional mean and variance models can be found  from the conditional mean and dispersion models on the noise variables, see Subsection \ref{Mean_Variance_Models}. The models for $\mu=E(Y|Z_1,Z_2)$ and $\log(\phi)$, whose parameter estimates are displayed in Table \ref{Tab_Examp_Mean_Disp}, are given by the equations (\ref{eq5_1}) and (\ref{eq5_2}), respectively.

\begin{eqnarray}
\label{eq5_1}
E(Y|Z_1,Z_2) & = & 488.961x_1 + 432.21x_2 + 574.124x_3 + 174.216x_1x_3Z_1 + \nonumber \\
            &    & (56.621x_1 + 35.904x_2 + 79.146x_3)Z_2
\end{eqnarray}

\begin{equation}
\label{eq5_2}
\log(\phi) = 6.9984x_1 + 5.94x_2 + 7.325x_3 - 7.9662x_2x_3 
\end{equation}

Thus, considering $Z_1$ and $Z_2$ independent random variables with $E(Z_i)=\mu_i$ and $Var(Z_i)=\sigma_{i}^{2}$, for $i=1.2$ , the unconditional mean model is obtained as $E(Y)=E[E(Y|Z_1,Z_2)]$ and shown in the equation (\ref{eq5_3}).

\begin{eqnarray}
\label{eq5_3}
E(Y) & = & 488.961x_1 + 432.21x_2 + 574.124x_3 + 174.216x_1x_3\mu_1 + \nonumber \\
            &    & (56.621x_1 + 35.904x_2 + 79.146x_3)\mu_2.
\end{eqnarray}

The unconditional variance model is obtained as $Var(Y)=E[Var(Y|Z_1,Z_2)] + Var[E(Y|Z_1,Z_2)]$. Since $Var(Y|Z_1,Z_2)=\phi V(\mu)$ and as we are assuming that $V(\mu)=1$ and we know the model for $\log(\phi)$, shown in equation (\ref{eq5_2}), we have that $Var(Y|Z_1,Z_2)=\exp\{6.9984x_1 + 5.94x_2 + 7.325x_3 - 7.9662x_2x_3\}$. Thus, as $Var[E(Y|Z_1,Z_2)]=(174,216x_1x_3)^{2}\sigma_{1}^{2} + (56,621x_1 + 35,904x_2 + 79,146x_3)^{2}\sigma_{2}^{2}$ and $E[Var(Y|Z_1,Z_2)]=\exp\{6.9984x_1 + 5.94x_2 + 7.325x_3 - 7.9662x_2x_3\}$ we get that

{\scalefont{0.9}
\begin{eqnarray}
\label{eq5_4}
Var(Y) & = & \exp\{6.9984x_1 + 5.94x_2 + 7.325x_3 - 7.9662x_2x_3\} + \nonumber \\
            &    & (174.216x_1x_3)^{2}\sigma_{1}^{2} + (56.621x_1 + 35.904x_2 + 79.146x_3)^{2}\sigma_{2}^{2}. 
\end{eqnarray}
}

Note that, knowing the unconditional mean and variance models and setting values for $\mu_1$, $\mu_2$, $\sigma^2_1$ and $\sigma^2_2$, we could consider a robust design problem for which we are interested in an optimization problem, which minimizes the variance subject to the restriction that the mean is equal to 530 ml, that the mixture condition is satisfied and also subject to restrictions on the mixture variables, see Subsection \ref{Mean_Variance_Models} and \cite{GranatoCaladoPinto} for more details. 

\section{Conclusion}
\label{Conclusion}

In this article, we have shown how the joint modeling of mean and dispersion can be used to model variability in experiments involving mixtures. A methodology for variable selection in the JMMD applied to mixture experiments was introduced. The variable selection process proved to be efficient in obtaining the mean and dispersion models. In addition, our methodology also provides a way to verify constant dispersion. 

The variable selection process in JMMD, presented in Subsection \ref{Variable_Selection} and exemplified in Section \ref{Application}, allows us to quickly and effectively find the mean and dispersion models. The mean and variance models can be obtained directly from these models. When there are noise variables influencing the mixture composition, the models found for mean and variance are conditioned to these variables. The unconditional models can be found using expressions for unconditional mean and variance shown in Subsection  \ref{Mean_Variance_Models}. 

The variable selection procedure uses the same iterative process of the JMMD, see Subsection \ref{Estimation}, which makes the search for the optimal mean model sequential, that is, in each iteration of the process, the adjustment of the model improves. The dispersion model, due to the fact that its response is the deviance from the mean model, varies for each iteration of the procedure. 

When we use the JMMD, which belongs to the class of generalized linear models, other distributions, that not only the Gaussian distribution, could be considered for the mean model, for instance, distributions for counts or proportions. However, in robust design problems the complexity of the optimization problem can increase, see Subsection \ref{Mean_Variance_Models}.

The bread-making problem, considered in Section \ref{Application}, was also analyzed by \cite{NaesFaergestadCornell} and \cite{GranatoCaladoPinto}. The first considered only the modeling of the mean. The second, considering the presence of noise variables, used the delta method to model both mean and variance. In both approaches, classical regression models were considered, i.e., the variance function was $V(\mu)=1$ and the dispersion parameter $\phi$ was considered as constant. However, using the variable selection procedure for the JMMD, we found that for this data set the dispersion parameter is not a constant, see Table \ref{Steps_Procedure} - iteration 2. In classical regression models, constant dispersion means that the errors in the regression model are homoscedastic, i.e., they have the same variance. Thus, non-constant dispersion means heteroscedasticity. The existence of heteroscedasticity is a major concern in regression analysis and analysis of variance, because it invalidates statistical tests of significance that assume that all modeling errors have the same variance, making it much more likely that a regression model will declare that a term in the model is statistically significant, when in fact it is not. For more details on heteroscedasticity in classical regression models see, for example, \cite{GujaratiPorter}, p. 365. In this way, due to heteroscedasticity, the models found by \cite{NaesFaergestadCornell} and \cite{GranatoCaladoPinto} may have been seriously affected by the presence of non-constant dispersion and, therefore, the results of the analyzes obtained by them may be misleading. In the JMMD approach, non-constant dispersion is not a problem. The JMMD allows the construction of a statistical model for the dispersion and provides an efficient mechanism for obtaining a priori weights for the mean model. The model for the response variance can be obtained from the dispersion model and the variance function. 

It is also worth mentioning that in the approach using the JMMD, if the process presents variability, the mean and variance models can be obtained even without the presence of noise variables. Thus, the approach proposed in this article is more general to model the variability in experiments involving mixtures and, in the case of non-constant dispersion, provides more reliable results.

In this way, we can suggest the use of regression models in the mixture experiments as follows: i) if there are no noise variables and if we verify that the dispersion parameter is constant, then we will only have the mean modeling and we can use classical regression model or generalized linear model, depending on the type of distribution to be considered for the response variable; ii) if we verify that the dispersion parameter is constant, but if there are noise variables in the process, then we can model the mean and variance using the delta method, considering classical regression model or GLM to find the mean model; iii) if we verify that the dispersion parameter is not constant and if there are no noise variables in the process, then the mean and variance models can be obtained using the JMMD; iv) if in addition to the non-constant dispersion there are also noise variables, then the mean and variance models can be obtained as follows. Initially, the mean and dispersion models, conditional on the noise variables, are obtained using the JMMD; later, the mean and variance models, unconditional on the noise variables, are found according to what was exposed in Subsection \ref{Mean_Variance_Models}.

The results obtained in this article through the application of joint modeling of mean and dispersion in mixture experiments were very encouraging, indicating that the theory promises to be of great utility in modeling and conducting industrial experiments involving mixtures. A function in R software that performs the process of searching for the optimal joint model is available in \cite{PereiraPinto}.

\begin{acks}[Acknowledgments]

The authors would like to thank the anonymous referees and the Editor for their constructive comments that improved the quality of this paper.
\end{acks}

\begin{funding}
This work was supported by Research Support Foundation of the State of Minas Gerais (FAPEMIG) under Grant Number APQ-03365-18.

\end{funding}


 {}

\end{document}